\documentstyle[12pt,epsfig]{article}

\hoffset=- 0.5 truecm
\def\R{ {\rm R \kern -.31cm I \kern .15cm}}
\def\C{ {\rm C \kern -.15cm \vrule width.5pt \kern .12cm}}
\def\Z{ {\rm Z \kern -.27cm \angle \kern .02cm}}
\def\N{ {\rm N \kern -.26cm \vrule width.4pt \kern .10cm}}
\def\1{{\rm 1\mskip-4.5mu l} }

\newcommand{\be}{\begin{equation}}
\newcommand{\ee}{\end{equation}}

\def\noi{\noindent}

\def\lsim{\raise0.3ex\hbox{$<$\kern-0.75em\raise-1.1ex\hbox{$\sim$}}}
\def\gsim{\raise0.3ex\hbox{$>$\kern-0.75em\raise-1.1ex\hbox{$\sim$}}}
 \def\cite#1{[\ref{#1}]}

\begin{document}
\baselineskip=20 pt

\begin{center}
{\Large \bf Pressure and Maxwell Tensor} \par \vskip 3 truemm
{\Large \bf in a Coulomb Fluid} \par \vskip 1 truecm
{\bf B. Jancovici}$^1$
\end{center}
\vskip 2 truecm
\begin{abstract}
The pressure in a classical Coulomb fluid at equilibrium is obtained 
from the Maxwell tensor at some point inside the fluid, by a suitable
statistical average. For fluids in an Euclidean space, this is a
fresh look on known results. But, for fluids in a curved space, a case
which is of some interest, the unambiguous results from the Maxwell
tensor approach have not been obtained by other methods. 
\end{abstract}

\vspace{1 truecm}

\vskip 3 truecm

\noindent LPT Orsay 99-103 \par
\noindent December 1999 \par
\vfill
\hbox to 4 truecm{\hrulefill}
\noi $^1$ Laboratoire de Physique Th\'eorique, Universit\'e de Paris-Sud,
B\^atiment 210, 91405 Orsay, France (Unit\'e Mixte de Recherche 
n$^{\circ}$ 8627 - CNRS). \par
\noi E-mail: Bernard.Jancovici@th.u-psud.fr

\newpage
\pagestyle{plain}
\noi {\bf 1. INTRODUCTION} \\

\noi The aim of the present paper is to revisit the concept of pressure
for a Coulomb fluid, i.e. a fluid made of
particles interacting through Coulomb's law (electrolyte,
plasma,...). We consider only fluids in thermodynamic equilibrium, and
assume that classical (i.e. non-quantum) statistical mechanics is
applicable. \par

Pressure is often defined as the force per unit area that a fluid exerts
on the walls of a (large) vessel containing it. However, pressure may
also be defined without reference to any wall. One has to {\it imagine} 
some {\it immaterial} plane surface across the fluid, and pressure is
the force per unit area with which the fluid lying on one side of this
surface pushes on the fluid lying on the  other side. Both definitions
agree with each other. \par

From a microscopic point of view, the force between two parts of the
fluid is usually described in terms of the interactions between the
molecules. Two molecules at a distance $r$ from each other are supposed
to interact through some potential. This is the standard approach, which
is briefly recalled in Section 2. \par 

In the case of electromagnetism, Maxwell, following Faraday, came
to a different point of view: the forces between two charged objects are
mediated by fields. At a given point of space, even in vacuum
outside the charges, there is a stress tensor (the Maxwell stress
tensor), a local quantity defined in terms of the fields at that point,
similar to the stress tensor within some elastic medium. In this
picture, every region of ``empty'' space exerts forces on the regions
ajacent to it. In Section 3, it is shown how the pressure at some point
inside a Coulomb fluid can be defined and computed from the Maxwell
tensor at that point by a suitable statistical average, with an
appropriate prescription for obtaining a finite result. This result
agrees with the standard one. \par

Section 3 ``extends'' the above ideas to the case of two-dimensional
models. \par 

Section 4 discusses the case of Coulomb fluids living in a
curved space. In this case, the Maxwell tensor approach will be shown to
be especially appropriate. \newpage

\noi {\bf 2. A SUMMARY OF THE STANDARD APPROACH} \\

In the simple case of a fluid made of one species of particles,  
with a pair interaction $v(r)$ depending only on the distance $r$,
the pressure $P$ is found to be, in the thermodynamic limit,
$$P=nkT-{1\over 6}n^2\int {dv\over dr}rg(r)d{\bf r} \eqno(2.1)$$

\noi where $n$ is the number density (number of particles per unit
volume), $k$ is Boltzmann's constant, $T$ is the temperature, and $g(r)$ is
the pair distribution function. In (2.1), $nkT$ is the ideal gas part of
the pressure (related to the momentum carried by the particles), while
the following term, due to the interactions, is called the excess
pressure $P_{ex}$. The same equation (2.1) is obtained by looking either 
at the pressure on the walls$^{(1)}$ or at the pressure in the bulk
fluid$^{(2)}$. \par

Real Coulomb fluids are made of several species of particles (for
instance, in an electrolyte, positive and negative ions, plus the solvent
molecules). In a classical model, some short-range non-Coulombic
interaction must be introduced, for avoiding the collapse on each other
of particles of opposite sign. Here, for simplicity, we rather consider
only a simplified model, 
the one-component plasma (OCP)$^{(3)}$: identical point-particles of one
sign, each of them carrying an electric charge $q$, embedded in a uniform
background of opposite sign which ensures overall neutrality. Only the
Coulomb interaction is retained, thus the interaction is $v(r)=q^2/r$
and $(dv/dr)r=-q^2/r$. Due to the background, the
average charge in a volume element $d{\bf r}$ at a distance $r$ of a
given particle is $qn[g(r)-1]$ rather than $qng(r)$, and, in the case
of the OCP, equation (1) is to be replaced by
$$P=nkT+{q^2n^2\over 6}\int {1\over r}h(r)d{\bf r} \eqno(2.2)$$

\noi where $h(r)=g(r)-1$ is the pair correlation function. It
may be noted that $P_{ex}$ is one third of the potential energy
density.\footnote{For an OCP, there are several non-equivalent possible
definitions of the pressure. The pressure (2.2) is the thermal pressure,
in the sense of Choquard et al.$^{(4)}$} \par

The above standard approach is based on the assumption of an
interaction-at-distance $q^2/r$. In the next Section, it will be shown
how (2.2) can be derived by using the Maxwell tensor. \newpage

\noi {\bf 3. THE MAXWELL TENSOR APPROACH} \\

If only electrostatic interactions are retained (magnetic effects
are neglected), the Maxwell tensor is$^{(5)}$
$$T_{\alpha \beta }={1\over 4\pi }(E_{\alpha }E_{\beta }-
{1\over 2}{\bf E}\cdot {\bf E}\,\delta_{\alpha \beta }) \eqno(3.1)$$

\noi In (3.1), the Greek indices label the three Cartesian axes $(x,y,z)$. 
$T_{\alpha \beta }$ is the $\alpha $ component of the force per unit
area transmitted, across a plane normal to the $\beta $ axis, to the fluid
lying on the negative side of this plane. Thus, choosing for $\beta $
any axis, say the $x$ axis, one obtains for the excess pressure, which
is a force along that axis, 
$$P_{ex}=-<T_{xx}>=-{1\over 8\pi }<E_{x}^{\,2}-E_{y}^{\,2}-E_{z}^{\,2}>
\eqno(3.2)$$

\noi where $<\ldots >$ denotes a statistical average on all particle
configurations (the electric field at some point is a function of the
particle configuration). Our task is to evaluate the statistical average
(3.2) at some point inside the fluid, say at the origin. \par

Let $\rho^{(2)}(r_{12})$ be the statistical average of the product
microscopic charge density at ${\bf r}_1$ times microscopic charge
density at ${\bf r}_2$ (${\bf r}_{12}={\bf r}_2-{\bf r}_1$). From (3.2),  
$$P_{ex}=-{1\over 8\pi }\int d{\bf r}_1d{\bf r}_2\frac{x_1x_2-y_1y_2-z_1z_2}
{r_{1}^{\,3}r_{2}^{\,3}}\rho^{(2)}(r_{12}) \eqno(3.3)$$

\noi In the present case of an OCP,
$$\rho^{(2)}(r_{12})=q^2[n\delta({\bf r}_{12})+n^2h(r_{12})] \eqno(3.4)$$

\noi Using (3.4) in (3.3) gives to $P_{ex}$ two contributions $P_{self}$
and $P_{nonself}$ involving the $\delta $ part and the $h $ part of
(3.4), respectively. $P_{nonself}$  gives no difficulty. This is a
convergent integral (indeed, $h$ is -1 at small $r_{12}$ because the
particles strongly repel each other, and $h$ has a fast decay at large
$r_{12}$ because remote particles are uncorrelated). Because of the
rotational symmetry around the origin, it can be rewritten as
$$P_{nonself}={q^2n^2\over 24\pi }\int d{\bf r}_1d{\bf r}_2\frac{{\bf
r}_1\cdot {\bf r}_2}{r_{1}^{\,3}r_{2}^{\,3}}h(r_{12}) \eqno(3.5)$$  

\noi But 
$$P_{self}=-{nq^2\over 8\pi }\int d{\bf r}\frac{x^2-y^2-z^2}{r^6}\eqno(3.6)$$

\noi diverges at small $r$.

The resolution of the difficulty is that the force that each particle exerts
on itself should not be taken into account. Thus, the integral in (3.6)
must be regularized by the prescription that no particle sits on the
$x=0$ plane on which we have chosen to compute the pressure force. This
prescription can be enforced by removing from the integration domains
a thin slab $-\varepsilon <x<\varepsilon $ and taking the limit 
$\varepsilon \rightarrow 0$ at the end. This prescription does not
change the convergent integral (3.5). But it means that the self part
(3.6) must be defined, in cylindrical coordinates $(x,\rho)$, as
$$P_{self}=-\frac{nq^2}{8\pi}\lim_{\varepsilon \rightarrow 0}\int_{|x|>\varepsilon} dx\int_{0}^{\infty} 
2\pi d\rho \rho \frac{x^2-\rho^2}{(x^2+\rho^2)^3} \eqno(3.7)$$

\noi Since the integral on $\rho$, performed first, is found to
vanish, the result is $P_{self}=0$.

As to $P_{nonself}$, (3.5) can be easily computed by taking as integration
variables ${\bf r}_1$ and ${\bf r}_{12}$, and performing the integral on 
${\bf r}_1$ first with the result $4\pi /r_{12}$. The final result is  
$$P_{ex}=P_{nonself}={q^2n^2\over 6}\int d{\bf r}_{12}{1\over r_{12}}h(r_{12})
\eqno(3.8)$$

\noi in agreement with the standard formula (2.2). \par

An alternative way of calculating $P_{self}$ will turn out to be more
appropriate for extensions which follow. (3.6) is split into the
contributions $P_0$ of $r<r_0$ and $P_1$ of $r>r_0$, where $r_0$ is the
radius of a small sphere centered at the origin. The prescription that
no particle sits on the plane $x=0$ does not change the convergent part
$P_1$, which can be computed, using the rotational symmetry, as
$$P_1=\frac{nq^2}{24\pi }\int_{r>r_0} \frac{d{\bf r}}{r^4}
=\frac{nq^2}{6r_0} \eqno(3.9)$$

\noi It is only in $P_0$ that the rotational symmetry is broken by the
prescription $|x|>\varepsilon $, which gives
$$P_0=-\frac{nq^2}{8\pi }\lim_{\varepsilon \rightarrow
0}\int_{\varepsilon <|x|<r_0}dx\int_{0}^{\sqrt{r_{0}^{\,2}-x^2}}2\pi d\rho
\rho \frac{x^2-\rho^2}{(x^2+\rho^2)^3}=-\frac{nq^2}{6r_0} \eqno(3.10)$$

\noi Thus $P_{self}=P_0+P_1=0$, and (2.2) is retrieved. \newpage
 
\noi {\bf 4. TWO-DIMENSIONAL MODELS} \\

Two-dimensional models of Coulomb fluids are of interest for at
least two reasons. First,some of these models are physically relevant.
Second, exact results are available. The two-dimensional case has
special features which require the present separate discussion.\par

In two dimensions, the Coulomb interaction (as defined through the
Poisson equation) between two charges $q$ and $q'$ is $-qq'\ln (r/L)$,
where $L$ is some irrelevant length. Since this interaction diverges at
$r=0$ only mildly, in addition to the OCP it is also possible to
consider a two-component plasma (TCP), made of positive and negative
point-particles of respective charges $q$ and $-q$, without any
additional short-range repulsion (which is stable provided that the
coupling constant $\Gamma=q^2/kT$ be smaller than 2). \par

For the OCP, the two-dimensional analog of (2.1), with the background
taken into account, is
$$P=nkT-{1\over 4}n^2\int {dv\over dr}rh(r)d{\bf r} \eqno(4.1)$$

\noi Now $v(r)=-q^2\ln (r/L)$ and (4.1) becomes
$$P=nkT+{1\over 4}n^2q^2\int h(r)d{\bf r} \eqno(4.2)$$

\noi Perfect screening, present in a conductor, says that
$$n\int h(r)d{\bf r}=-1 \eqno(4.3)$$

\noi (this means that the polarization cloud around a particle of charge
$q$ carries the opposite charge $-q$). Using (4.3) in (4.2) gives the
simple exact equation of state~$^{(6,7)}$
$$P=n(kT-\frac{q^2}{4}) \eqno(4.4)$$

\noi Now, $P_{ex}=-nq^2/4$ is no longer related to the potential energy
density.  \par 

We now turn to the Maxwell tensor approach. In two dimensions, the
Maxwell tensor is
$$T_{\alpha \beta }={1\over 2\pi }(E_{\alpha }E_{\beta }-
{1\over 2}{\bf E}\cdot {\bf E}\,\delta_{\alpha \beta }) \eqno(4.5)$$

\noi with Greek indices now labeling two Cartesian axes $(x,y)$. (3.3) is
replaced by
$$P_{ex}=-{1\over 4\pi }\int d{\bf r}_1d{\bf r}_2\frac{x_1x_2-y_1y_2}
{r_{1}^{\,2}r_{2}^{\,2}}\rho^{(2)}(r_{12}) \eqno(4.6)$$

\noi where, for an OCP, (3.4) still holds. Now, although (4.6) still
converges for large values of $r_1$ and $r_2$ (because
$\rho^{(2)}(r_{12})$ has a fast decay as $r_{12}$ increases and 
its integral vanishes), separating it in self and
nonself parts would generate terms separately diverging at infinity.
Here, it is more appropriate to split (4.6) in another way, similar to
what has been done at the end of Section 3. Namely, in (4.6), one
separates the contribution $P_0$ of the integration domain
$(r_1,r_2<r_0)$ and the rest $P_{ex}-P_0$. This rest is a convergent
integral and, by rotational symmetry, it vanishes. One is left with
$P_0$ which can be split into its self and nonself parts, with now a
nonself part which is convergent and also vanishes by rotational symmetry. 
Finally, the self part has to be defined in the same way as (3.10), and     
$$P_0=-\frac{nq^2}{4\pi }\lim_{\varepsilon \rightarrow
0}\int_{\varepsilon <|x|<r_0}dx\int_{-\sqrt{r_{0}^{\,2}-x^2}}^{\sqrt{r_{0}^{\,2}-x^2}}dy\frac{x^2-y^2}{(x^2+y^2)^2}=-\frac{nq^2}{4} \eqno(4.7)$$

\noi Thus
$$P_{ex}=P_0=-\frac{nq^2}{4} \eqno(4.8)$$

\noi in agreement with (4.4). \par

Similar considerations hold for the TCP, as long as $\Gamma <2$, and the
equation of state again is (4.4), where now $n$ is the total number
density of the particles. \par

The two-dimensional OCP can also be obtained as a limit of the
$\nu$-dimensional one, as explained in Appendix A. \\

\noi {\bf 5.CURVED SPACES} \\

The statistical mechanics of a Coulomb fluid living in a curved
space is of interest for at least two reasons. First, for doing
numerical simulations (necessarily on a finite system) without having to
deal with boundary effects, a clever method has been to confine the
system on the surface of a sphere (in the two-dimensional case)$^{(8)}$
or an hypersphere (in the three-dimensional case)$^{(9,10)}$. Second, for 
two-dimensional Coulomb fluids on a surface of constant negative
curvature (pseudosphere) $^{(11)}$, it is possible to go to the limit of
an infinite system while keeping a finite curvature, thus to look at the
properties of a curved infinite system (something which cannot be done
for a sphere or hypersphere). \par

The present paper actually arose from the question: How to define the
pressure of a Coulomb fluid in a curved space, away frow any wall? A
formula like (2.1) is based on the interaction-at-distance picture: the
force acting on the fluid lying on one side of some immaterial plane is
the sum of elementary forces acting on each molecule. This picture
cannot be generalized to the case of a curved space, because there is no  
straightforward way of summing forces (vectors) applied at different
points of space. Thus, the Maxwell tensor picture seems to be the only
possible one, defining the pressure at a given point of space as
a local quantity depending only on the electric field at this
point. \par

Three kinds of Coulomb fluids will be considered: The three-dimensional
OCP on a hypersphere, the two-dimensional OCP or TCP on a sphere, the
two-dimensional OCP or TCP on a pseudosphere. \\

\noi {\bf 5.1.OCP on a Hypersphere} \\ 

The hypersphere is the four-dimensional analog of the usual
sphere. We consider an OCP living on the three-dimensional ``surface''
$S_3$ of a hypersphere of radius $R$. On $S_3$, the geodesic distance
between two points is $R\psi $, where $\psi \in [0,\pi ]$ is the angular
distance between these points, as seen from the center of the
hypersphere. The volume element between two concentric spheres of radii
$R\psi $ and $R(\psi +d\psi )$ is $dV=4\pi R^3\sin^2\psi d\psi $. The
total volume of $S_3$ is $V=2\pi^2R^3$. \par
 
Since $S_3$ is a compact manifold without boundary, electric potentials
and fields can be 
defined only if the total charge is zero. In particular, the electric
field created by one point charge cannot be defined. For overcoming this
difficulty, one can consider the OCP as a collection of pseudocharges
$^{(9)}$: a pseudocharge is defined as a point charge $q$ plus a uniform
background of total charge $-q$. At a point $M$ located at a geodesic
distance $R\psi $ from a pseudocharge located at $M_0$, the electric
potential created by the pseudocharge is
$$\Phi =\frac{q}{\pi R}\left( (\pi -\psi )\hbox{ctn}\,\psi
-\frac{1}{2}\right) +V_0 \eqno(5.1)$$

\noi where $V_0$ is an arbitrary constant. The corresponding electric
field at $M$ is
$${\bf E}=\frac{q}{\pi R^2}\left(\hbox{ctn}\,\psi +\frac{\pi
-\psi}{\sin^2\psi}\right){\bf t} \eqno(5.2)$$

\noi where ${\bf t}$ is the unit vector tangent to the geodesic $MM_0$ at
$M$. From (5.1), the interaction energy between two pseudocharges $i$
and $j$ at a geodesic distance $R\psi_{ij} $ of each other is found to be
$$\phi (\psi_{ij})=\frac{q^2}{\pi R}\left((\pi -\psi_{ij})
\hbox{ctn}\,\psi_{ij}-\frac{1}{2}\right)  \eqno(5.3)$$

\noi independent of $V_0$. \par

The excess pressure is given by (3.2) where the electric field can be
written as ${\bf E}=\sum {\bf E}_i$, with ${\bf E}_i$ the field created
by the \textit{i}-th pseudocharge.  As above, (3.2) can be split into a
self part $P_{self} $ (made of ${\bf E}_i{\bf E}_i $ terms) and a nonself
part $P_{nonself} $ (made of ${\bf E}_i{\bf E}_j\,\,(i\not=j) $
terms). \par 

Because of the rotational symmetry, $P_{nonself}$ can be written as
$$P_{nonself}=\frac{1}{24\pi}\langle \sum_{i\not=j}{\bf E}_i\cdot
{\bf E}_j\rangle=\frac{1}{3}u_{nonself} \eqno(5.4)$$
   
\noi where $u_{nonself}$ is the nonself part of the potential energy
density, which can be reexpressed in terms of the interaction $\phi $
rather than in terms of fields, by the usual integration by parts, as
$$u_{nonself}=\frac{1}{2\pi^2R^3}\langle \sum_{i<j}\phi\ (\psi_{ij})
\rangle =\frac{n^2}{2}\int \phi (\psi )h(\psi )dV \eqno(5.5)$$

\noi where one can use the pair correlation function $h$ rather than the 
pair distribution function $g=h+1$ since $\int \phi dV=0$.
 
As to $P_{self}$, it is a divergent integral which however can be made
finite by adapting what has been done at the end of Section 3, namely 
splitting it into the  contribution $P_0 $ of geodesic distances $R\psi 
<R\psi_0 $, and the finite contribution $P_1 $ of $R\psi >R\psi_0 $ for
which rotational symmetry can be used. In terms of $E(\psi) $ given by
(5.2),
$$P_1=\frac{n}{24\pi }\int_{\psi_0}^{\pi}E^2(\psi )4\pi R^3\sin^2 \psi
 d\psi \eqno(5.6)$$   

\noi Evaluating the integral in (5.6) as $\psi_0 \rightarrow 0 $ gives 
$$P_1=\frac{nq^2}{6R}\left( \frac{1}{\psi_0}-\frac{3}{2\pi}+O(\psi_0)
\right) \eqno(5.7)$$
 
\noi On the other hand, for $P_0$, the curvature effects become
negligible as $\psi_0 \rightarrow 0 $ (more precisely, as shown in
Appendix B, they are $O(\psi_0)$) and the Euclidean prescription (3.10)
can be used, with $r_0=R\psi_0$, giving 
$$P_0=-\frac{nq^2}{6R\psi_0}+O(\psi_0) \eqno(5.8)$$

Finally, the total pressure is
$$P=nkT+\frac{n^2}{6}\int \phi (\psi )h(\psi )dV-\frac{nq^2}{4\pi R}
\eqno(5.9)$$ 

\noi This is the generalization of (2.2) to the case of a hypersphere.

The present evaluation of the pressure makes no explicit use of the
self-energy of a pseudoparticle. This is an important remark. Indeed,
another possible definition of the pressure would be minus the
derivative of the free energy with respect to the volume. But, for
evaluating the free energy, it is necessary to define properly the zero
of energy for a system of pseudoparticles, and this necessarily involves 
some heuristic convention about the self-energy of a pseudoparticle. In
ref.9. a reasonable convention gave $-3q^2/4\pi R$ for this self-energy,
and the corresponding pressure is identical with (5.9). This pressure
does obey the usual relation 
$$P_{ex}=\frac{1}{3}u \eqno(5.10)$$

\noi where $u$ is the total potential energy density defined with the
above convention. However, in ref.10, another reasonable convention
(giving a faster approach to the thermodynamic limit as $R \rightarrow
\infty $) has been used, with additional terms of order higher than
$1/R$, and the pressure derived from the corresponding free energy no
longer agrees with (5.9). The definition of the pressure in terms of the
Maxwell tensor is free from this arbitrariness. \par
\newpage
\noi {\bf 5.2.OCP or TCP on a Sphere} \\

The above considerations can be easily adapted to the (simpler) case of 
two-dimensional Coulomb systems living on the surface of a sphere. Now,
the electric potential created by a pseudocharge is
$$\Phi=-q\ln \sin (\psi /2)+V_0 \eqno(5.11)$$ 

\noi $P_{nonself}$ and $P_1$,  convergent integrals, vanish because of
the rotational symmetry. One is left with $P_0$, for which the curvature
effects are negligible and (4.7) holds with the same result (4.8) as in
the case of a plane system. This result $P_{ex}=-nq^2/4$ holds for an
OCP, and also for a TCP when $\Gamma <2$. \par

Here too, the free energy depends on an arbitrary convention about the
zero of energy. In ref.8, this convention was implicitly made by
the way in which (5.11) was used together with the choice $V_0=-q\ln
(2R/L)$, with $R$ the radius of the sphere. It is only thanks to this
convention that the corresponding free energy has a derivative with
respect to the sphere area which correctly gives the equation of state
(4.4). \\

\noi {\bf 5.3.OCP or TCP on a Pseudosphere} \\

Recently, two-dimensional Coulomb systems living on a surface of
constant negative curvature (a pseudosphere) were studied.$^{(11)}$
Since, unlike a sphere, a pseudosphere infinite, one has the
interesting possibility of considering systems which are both infinite
and curved. \par 

Let $a$ be a length such that the Gaussian curvature of the pseudosphere
is $-1/a^2$ (instead of $1/R^2$ on a sphere). Now, the electric
potential and field created by a single point charge $q$ exist. At a
geodesic distance $s$ from this charge, the electric potential is
$$\Phi=-q\ln \tanh \frac{s}{2a} \eqno(5.12)$$
     
\noi where the possible additive constant has been fixed by the
condition that this potential vanishes at infinity ($s \rightarrow
\infty$). The electric field is    
$${\bf E}=\frac{q}{a\sinh (s/a)}{\bf t} \eqno(5.13)$$

\noi where ${\bf t}$ is the unit vector tangent to the geodesic. \par 

The pressure can be obtained from the Maxwell tensor just as in the case
of a sphere, with the same result (4.4), i.e. $P_{ex}=-nq^2/4$. This
pressure holds not only in the thermodynamic limit, but also at the
center of a finite disk. \par

The above result for the pressure calls for some discussion. On a
pseudosphere, when the size of a large domain increases, its perimeter
grows as fast as its area. As a consequence, there is no uniquely
defined thermodynamic limit for the free energy per particle
(this limit depends on the shape of the domain and on the boundary
conditions). A bulk pressure cannot be defined by deriving the free
energy with respect to the area. In ref.11, a bulk ``pressure'' $p$ was
defined by its virial expansion with the prescription that the
thermodynamic limit of each virial coefficient $B_k$ (which seems to
exist on a pseudosphere) has to be computed before the virial series in
powers of the density $n$ is summed. It is now apparent that this $p$ is
not identical with the pressure $P$ obtained from the Maxwell tensor in
the form (4.4). We now believe that the correct pressure is $P$, while
$p$ only is a mathematical quantity (seemingly with some interesting
properties).  \par

Nevertheless an important result of ref.11 is true: there is at least
one thermodynamic quantity, the bulk energy per particle, which has a
series expansion in integer powers of the density, in contrast to the
case of a plane system in which the energy per particle is singular at
zero density. \\

\noi {\bf 5.4.Why no Trace Anomaly?} \\

Some time ago, it has been remarked that conducting 
Coulomb systems are critical-like at any temperature$^{(12,13)}$ in some
sense: they have long-range {\it electric potential and field}
correlations, and the free energy of a two-dimensional
Coulomb system with an electric potential $\phi$ has logarithmic
finite-size corrections similar to the ones which occur$^{(14)}$ in a
critical system described by a conformal-invariant field theory with a 
field $\phi$. In ref.14, it was shown that, for a critical system,
this logarithmic correction to the free energy is associated to 
a trace anomaly of the stress tensor, proportional to the Gaussian
curvature of the surface on which the system lives. Thus, at first
sight, one might expect that the pressure of a two-dimensional Coulomb
system (which is minus one half of the expectation value of the trace of
the Maxwell tensor, with a suitably defined self part) would have a term
$O(1/R^2)$ on a sphere and $O(1/a^2)$ on a pseudosphere. Yet, such terms
are not present in (4.4). Why? \par 

Actually, for a Coulomb system made of $N$ particles on a sphere of
radius $R$, thus with a density $n=N/4\pi R^2$, the free energy $F$ 
has a term $(kT/6)\ln N$. The pressure is (with a suitable definition of
the zero of energy) a partial derivative at constant $N$:
$P=-\left(\partial F/\partial (4\pi R^2)\right)_N$.
Thus the $\ln N$ term in $F$ gives no contribution to the pressure, in
agreement with (4.4). However, in field theory, some ultraviolet cutoff
(a length $\eta$) has to be introduced and the trace of the stress
tensor has an expectation value $<\Theta>$ related to 
$\left(\partial F/\partial (4\pi R^2)\right)_{\eta}$ with now a
derivative taken at constant cutoff. When the Coulomb system is
described in terms of a field theory, the role of the cutoff is played
by the microscopic scale $\eta =n^{-1/2}$. Thus, the
$\ln N=\ln (n4\pi R^2)$ term in $F$ is associated to a trace anomaly in
the field-theoretical $<\Theta>$, not in $P$. \par

A related statement is: If the expectation value of the trace of the
Maxwell tensor is computed with a field-theoretical measure (the
functional integral measure $\mathcal{D}\phi$), it has a trace anomaly.
This trace anomaly is not present when the measure is the particle
configuration space one $d{\bf r}_1d{\bf r}_2\ldots d{\bf r}_N$. \par

Similarly, the pressure (4.8) on a pseudosphere has no trace anomaly. \\

\noi {\bf 6.CONCLUSION} \\

The pressure in a Coulomb fluid has been defined as minus
the statistical average of a diagonal element, say $T_{xx}$ of the Maxwell
tensor. This definition leads to an ill-defined integral, which however
can be given a definite value by an appropriate prescription: the fluid
is supposed split into two regions separated by a thin empty slab normal
to the $x$-axis, $T_{xx}$ is computed at a point inside this slab, and the
limit of a slab of zero thickness is taken at the end. \par

For Coulomb fluids in an Euclidean space, this approach through the
Maxwell tensor is just a fresh look on well-known results. But, for
Coulomb systems in a curved space, we are not aware of any other way of
obtaining an unambiguous value for the pressure.

For simplicity, only point-particle systems without short-range forces
have been considered. But an extension to systems with hard cores seems
feasible. \\

\noi {\bf APPENDIX A. THE $\nu$-DIMENSIONAL OCP} \\

In this Appendix, the excess pressure of a $\nu$-dimensional OCP 
($\nu>2$) is related to its potential energy density. The dimension
$\nu$ is treated as a continuous variable, and the limit $\nu
\rightarrow 2$ is taken. \par 

The unit of charge is defined such that the electric field at a distance 
$r$ from a unit charge be $1/r^{\nu -1}$. Thus, the potential is
$1/(\nu -2)r^{\nu -2}$. The Maxwell tensor is
$$T_{\alpha \beta}=\frac{1}{S_{\nu -1}}(E_{\alpha }E_{\beta}
-\frac{1}{2} {\bf E}\cdot {\bf E}\,\delta_{\alpha \beta})
\eqno(\hbox{A}.1)$$

\noi where
$$S_{\nu -1}=\frac{2\pi^{\nu /2}}{\Gamma (\nu /2)} \eqno(\hbox{A}.2)$$

\noi is the area of the sphere of unit radius. \par

In terms of the Maxwell tensor $T$, the nonself part of the pressure is 
$$P_{nonself}=-\frac{1}{\nu }<\hbox{tr}\,T>_{nonself}=
\frac{\nu -2}{\nu }\frac{1}{2S_{\nu -1}}<{\bf E}^2>_{nonself}
\eqno(\hbox{A}.3)$$  

\noi where the nonself part of the electrostatic energy density is
$$\frac{1}{2S_{\nu -1}}<{\bf E}^2>_{nonself}=\frac{q^2n^2}{2S_{\nu -1}}
\int d{\bf r}_1 d{\bf r}_2 \frac{{\bf r}_1\cdot {\bf r}_2}{r_{1}^{\,\nu}
r_{2}^{\,\nu}}h(r_{12}) \eqno(\hbox{A}.4)$$

\noi Taking as integration variables ${\bf r}_1$ and ${\bf r}_{12}$,
and performing first the integral on ${\bf r}_1$, one finds, as
expected, the potential energy density
$$\frac{1}{2S_{\nu -1}}<{\bf E}^2>_{nonself}=\frac{q^2n^2}{2}\int 
d{\bf r}_{12}\frac{1}{(\nu -2)r_{12}^{\,\,\nu -2}}h(r_{12})
\eqno(\hbox{A}.5)$$

As to the self part of the pressure $P_{self}=-<T_{xx}>_{self}$, it must
be defined, like in (3.7), as 
$$P_{self}=-\frac{nq^2}{2S_{\nu -1}}\lim_{\varepsilon \rightarrow 0}
\int_{|x|>\varepsilon}dx\int_0^{\infty}S_{\nu -2}d\rho \rho^{\nu -2}
\frac{x^2-\rho^2}{(x^2+\rho^2)^{\nu}} \eqno(\hbox{A}.6)$$

\noi Here too, the integral on $\rho $, performed first, is found to
vanish, thus $P_{self}=0$. \par

Therefore, the final result for the excess pressure is
$$P_{ex}=\frac{q^2n^2}{2\nu }\int d{\bf r}_{12}\frac{1}{r_{12}^{\,\,\nu -2}}
h(r_{12}) \eqno(\hbox{A}.7)$$

In the limit $\nu \rightarrow 2$, using the perfect screening rule
(4.3), one retrieves
$$P_{ex}=-\frac{nq^2}{4} \eqno(\hbox{A}.8)$$ \\

\noi {\bf APPENDIX B. CURVATURE EFFECTS IN A SMALL SPHERE} \\

In this Appendix, (5.8) is derived. \par

In four-dimensional Euclidean space, with Cartesian coordinates
$(x,y,z,t)$, the surface $S_3$ of a hypersphere of radius $R$ centered
at the origin usually is parametrized by the hyperspherical coordinates 
$(u,v,w)$ related to the Cartesian ones by
$$x=\sin w\sin v\cos u,\,\,y=\sin w\sin v\sin u,\,\,z=\sin w\cos v,\,\,
t=\cos w$$
$$0\leq u\leq 2\pi,\,\,0\leq v\leq \pi,\,\,0\leq w\leq \pi 
\eqno(\hbox{B}.1)$$

\noi However, here, it is more convenient to parametrize $S_3$ by the
three independent variables $(x,y,z)$. We define $r=(x^2+y^2+z^2)^{1/2}$
and $\rho =(y^2+z^2)^{1/2}$. A useful relation is $r^2=R^2\sin^2 w$.
Using the Jacobian for the change of coordinates, one finds that
the volume element $dV=R^3\sin^2w\sin vdudvdw$ becomes
$$ dV=\frac{dxdydz}{\cos w} \eqno (\hbox{B}.2)$$

\noi The hypersphere pole $x=y=z=0$ will be called $O$. The geodesic
distance between $O$ and $(x,y,z)$ is $Rw$. \par

The part $P_0$ of the pressure can be evaluated at $O$, i.e. the
electric field in (3.2) is the one at $O$. $P_0$ is that part of
$P_{self}$ which is created by the pseudocharges located at a geodesic
distance from $O$ smaller than $R\psi_0$. The regularization
prescription is that there is no particle in a thin slab  
$|x|<\varepsilon$. The electric field $E(w){\bf t}$  created at $O$
by a pseudocharge at $(x,y,z)$ is given by (5.2) where $\psi =w$ and 
${\bf t}=(-x/r,-y/r,-z/r)$. Thus, with (B.2) taken into account,
the analog of (3.10) is
$$P_0=-\frac{n}{8\pi}\lim_{\varepsilon \rightarrow 0}
\int_{\varepsilon<|x|<r_0}\int_{0}^{\sqrt{r_{0}^{\,2}-x^2}}
\frac{2\pi d\rho \rho}{\cos w}\frac{x^2-\rho^2}{x^2+\rho^2}E^2(w)
\eqno(\hbox{B}.3)$$

\noi where $r_0=R\sin \psi_0$. An expansion in powers of $r/R$ gives
$$\frac{E^2(w)}{\cos w}=\frac{q^2}{(x^2+\rho^2)^2}[1+O(r^2/R^2)]
\eqno(\hbox{B}.4)$$   

\noi When the expansion (B.4) is used in (B.3), the leading term of
(B.3) is (3.10) and the next term (which gives a convergent integral
for which the $\varepsilon$ regularization is superfluous) is
$O(r_0/R^2)$. Using $r_0=R\sin \psi_0$, one does find 
$$P_0=-\frac{nq^2}{6R\psi_0}+O(\psi_0)$$
 
\noi i.e.(5.8). \\

\noi {\bf ACKNOWLEDGEMENTS} \\

The author has benefited from stimulating discussions with J.M.Caillol,
J.L.Cardy, A.Comtet, F.Cornu, A.Krzywicki, R.Omnes, and many others. \\  
 
\def\labelenumi{[\arabic{enumi}]}
\noindent {\bf REFERENCES}
\begin{enumerate}
\item\label{1r} J.-P.Hansen and I.R.McDonald, {\it Theory of Simple
Liquids} (Academic, London, 1986).
\item\label{2r} P.A.Egelstaff, {\it An Introduction to the Liquid State}
(Academic, London, 1967).
\item\label{3r} M.Baus and J.-P.Hansen, {\it Phys.Rep.} {\bf 59}:1
(1980).
\item\label{4r} Ph.Choquard, P.Favre, and Ch.Gruber, {\it J.Stat.Phys.}
{\bf 23}:405 (1980).
\item\label{5r} J.D.Jackson, {\it Classical Electrodynamics} (Wiley, New
York, 1962).
\item\label{6r} A.M.Salzberg and S.Prager, {\it J.Chem.Phys.} {\bf
38}:2587 (1963). 
\item\label{7r} E.H.Hauge and P.C.Hemmer, {\it Phys. Norv.} {\bf 5}:209
(1971). 
\item\label{8r} J.M.Caillol, {\it J.Physique-Lettres} {\bf 42}:L-245
(1981). 
\item\label{9r} J.M.Caillol and D.Levesque, {\it J.Chem.Phys.} {\bf
94}:597 (1991).
\item\label{10r} J.M.Caillol, {\it J.Chem.Phys.} {\bf 111}:6528 (1999).
\item\label{11r} B.Jancovici and G.T\'{e}llez, {\it J.Stat.Phys.}
{\bf 91}:953 (1998). 
\item\label{12r} B.Jancovici, G.Manificat, and C.Pisani, {\it
J.Stat.Phys.} {\bf 76}:307 (1998).
\item\label{13r} G.T\'{e}llez and P.J.Forrester, {\it J.Stat.Phys.} {\bf
97}: 489(1999).
\item\label{14r} J.L.Cardy and I.Peschel, {\it Nucl.Phys.B} {\bf 300}
[FS 22]:377 (1988). 

\end{enumerate}
  
\end{document}